\documentclass[aps, reprint, 10pt, superscriptaddress, twocolumn, prb]{revtex4-1}
\usepackage{hyperref}
\usepackage{float}
\usepackage{longtable}
\usepackage{mathtools}
\usepackage{physics}
\usepackage[utf8]{inputenc}
\usepackage[T1]{fontenc}
\usepackage{lmodern}
\usepackage{url}
\usepackage[table]{xcolor}
\usepackage{multirow}
\usepackage{siunitx}
\usepackage{booktabs}
\usepackage{verbatim} 

\usepackage{ulem}

\begin{document}

\title{Self-consistent layer-projected scissors operator for band structures of complex 2D van der Waals materials}

\author{Dario A. Leon}
\email{dario.alejandro.leon.valido@nmbu.no}
\affiliation{
 Department of Mechanical Engineering and Technology Management, \\ Norwegian University of Life Sciences, NO-1432 Ås, Norway
}
\affiliation{CAMD, Department of
Physics, Technical University of Denmark, DK - 2800 Kongens Lyngby,
Denmark}

\author{Mikael Kuisma}
\affiliation{CAMD, Department of
Physics, Technical University of Denmark, DK - 2800 Kongens Lyngby,
Denmark}
\author{Mikkel Ohm Sauer}
\affiliation{CAMD, Department of
Physics, Technical University of Denmark, DK - 2800 Kongens Lyngby,
Denmark}
\author{Jakob K. Svaneborg}
\affiliation{CAMD, Department of
Physics, Technical University of Denmark, DK - 2800 Kongens Lyngby,
Denmark}

\author{Mark K. Svendsen}
\affiliation{CAMD, Department of
Physics, Technical University of Denmark, DK - 2800 Kongens Lyngby,
Denmark}
\affiliation{NNF Quantum Computing Programme, Niels Bohr Institute,
University of Copenhagen, Universitetsparken 5, 2100 Copenhagen, Denmark}

\author{Stefano Americo} 
\affiliation{CNR-SCITEC, Via Golgi 19, 20133 Milano, Italy}
\affiliation{CAMD, Department of
Physics, Technical University of Denmark, DK - 2800 Kongens Lyngby,
Denmark} 

\author{Kristian Berland}
\affiliation{
 Department of Mechanical Engineering and Technology Management, \\ Norwegian University of Life Sciences, NO-1432 Ås, Norway
}
\author{Jens Jørgen Mortensen}
\affiliation{CAMD, Department of
Physics, Technical University of Denmark, DK - 2800 Kongens Lyngby,
Denmark}

\author{Kristian S. Thygesen} \email{thygesen@fysik.dtu.dk}
\affiliation{CAMD, Department of
Physics, Technical University of Denmark, DK - 2800 Kongens Lyngby,
Denmark}
\date{\today}

\begin{abstract}
We introduce a computationally efficient method to calculate the quasiparticle (QP) band structure of general van der Waals (vdW) heterostructures. A layer-projected scissors (LAPS) operator, which depends on the one-body density matrix, is added to the density functional theory (DFT) Hamiltonian. The LAPS operator corrects the band edges of the individual layers for self-energy effects (both intralayer and interlayer) and unphysical strain fields stemming from the use of model supercells. 
The LAPS operator is treated self-consistently whereby charge redistribution and interlayer hybridization occurring in response to the band energy corrections are properly accounted for. We present several examples illustrating both the qualitative and quantitative performance of the method, including MoS$_2$ films with up to 20 layers, bilayer MoS$_2$ in an electric field, lattice-matched MoS$_2$/WS$_2$ and MoSe$_2$/WSe$_2$ bilayers, and MoSe$_2$/WS$_2$ moir\'e structures. Our work opens the way for predictive modeling of electronic, optical, and topological properties of complex and experimentally relevant vdW materials.

\end{abstract}

\maketitle

\section{Introduction}
Two-dimensional (2D) materials, consisting of covalently bonded, chemically saturated atomic layers, display a wealth of interesting properties that differ from those of conventional bulk materials due to pronounced quantum confinement and many-body effects.  \cite{bhimanapati2015recent,mak2010atomically,wang2012electronics,radisavljevic2011single,gong2017discovery} Many properties of the simpler 2D materials, in particular the pristine monolayers, are today well understood. This is not least a result of powerful \emph{ab initio} calculations, which have provided quantitative insight into the fundamental quantum mechanical processes governing the properties of the 2D materials on the atomic and electronic scales. \cite{thygesen2017calculating,kruse2023two,qiu2013optical,druppel2017diversity,sodequist2024two} 

Compared to the pristine monolayers, complex 2D materials such as van der Waals (vdW) heterostructures or twisted homobilayers, remain poorly understood. An important reason for this is that \emph{ab initio} calculations, in particular beyond standard density functional theory (DFT), are currently out of reach for this class of materials. In fact, many interesting and experimentally relevant vdW materials contain hundreds or even thousands of atoms in a unit cell making them inaccessible for the diagrammatic many-body methods, which have been crucial for establishing our understanding of monolayers.   

It is well known that (semilocal) DFT tends to underestimate quasiparticle (QP) band gaps of semiconductors and insulators,\cite{godby1988self} while the many-body GW method\cite{hedin1965new,hybertsen1986electron} (and its vertex-corrected flavors\cite{shishkin2007accurate,schmidt2017simple}) yields band gaps in good agreement with experiments -- at least when the QP approximation applies. For homogeneous materials, the DFT band structure is often qualitatively correct and the band gap problem can be handled, e.g. using a "scissors operator" that shifts all unoccupied bands by a constant energy relative to the occupied bands. The situation is much more severe for interfaces and heterostructures, where DFT often yields qualitatively wrong band alignment compared to GW\cite{haastrup2018computational}.  In such cases, the DFT band gap problem cannot be easily handled.

While the electronic band structure is obviously important in its own right, it is also the basis for almost any perturbation theory as well as for the description and understanding of more advanced properties. An important example of the latter is the optical excitation spectrum. The low-energy optical spectrum of a 2D semiconductor is governed by strongly bound excitons.\cite{wirtz2006excitons,olsen2016simple} The nature of these excitons, particularly whether they are of intralayer or interlayer type in a given heterostructure, depends directly on the band alignment. Specifically, the lowest excitons will be of the interlayer (intralayer) type in systems with Type-II (Type-I) band alignment. Furthermore, the technologically relevant mixed interlayer excitons can form if the difference in interlayer and intralayer band gaps matches the exciton binding energy.\cite{deilmann2018interlayer,peimyoo2021electrical} An accurate description of the QP band structure is therefore foundational for the understanding and predictive modeling of the optical properties of vdW heterostructures.

When considering the band structure of a vdW heterostructure, it is important to realize that there is an intricate relation between the band alignment, the degree of interlayer hybridization, and the charge distribution at the interface. The latter will typically contain a dipole component that introduces a shift of the bands on the two sides of the interface. When the band alignment is shifted, the wave function hybridization across the interface is altered and thereby also the charge distribution. This relation implies that any attempt to correct the DFT band alignment must be performed self-consistently. In particular, it means that the widely used G$_0$W$_0$ approach (even when feasible) would not be reliable for a vdW heterostructure, because the DFT starting point could contain irremediable errors.

In this work, we introduce an efficient computational method for calculating the electronic band structure of a general vdW heterostructure. The method makes use of a layer-projected scissors (LAPS) operator that acts on the individual layers of the heterostructure and simulates the effect of a many-body electron self-energy. The LAPS operator is determined self-consistently in conjunction with the DFT effective potential, thus ensuring that interfacial dipoles and interlayer hybridization are consistent with the new band positions imposed by the LAPS operator. The method relies on a localized basis set to define the subspaces of the 2D layers and is implemented in the GPAW electronic structure code.

\section{Methods}
In this section, we introduce the LAPS operator and discuss the physical origin of the different contributions to the energy shifts entering its definition and how to obtain them in practice.

\subsection{The LAPS operator}
To obtain the QP band energies of a general vdW heterostructure, we perform a self-consistent DFT calculation (in this work we use the PBE xc-functional) with a layer-projected scissors (LAPS) operator included to mimic the effect of 
the electron self-energy in the individual layers. 
The LAPS operator is defined so as to reproduce the exact (or best target) band edge energies of the isolated monolayers, corrected for strain and image charge effects (see next section). 
Interlayer hybridization and charge transfer effects are thus described self-consistently at the DFT level but with the target valence and conduction band edges enforced on the individual layers via the LAPS operator. 

Formally, the LAPS operator is defined as
\begin{align}\label{eq:laps}
    \hat{\Sigma}_{\mathrm{LAPS}} = \sum_i^{ \mathrm{layers}}\sum_{n} \Big [\Delta \varepsilon_{\mathrm{v},i} \rho_{ni} + \Delta \varepsilon_{\mathrm{c},i} (1-\rho_{ni})\Big ]  |\rho_{ni}\rangle\langle \rho_{ni}|
\end{align}
where $\Delta \varepsilon_{\mathrm{v},i} $ and $\Delta \varepsilon_{\mathrm{c},i}$ represent the energy shift of the occupied and unoccupied states in the $i$th layer, respectively. 
$|\rho_{ni}\rangle$ should ideally represent a complete basis for the Hilbert space of the states of layer $i$, with occupation numbers $\rho_{ni}=1$ or $\rho_{ni}=0$. Defined in this way, $\hat \Sigma_{\mathrm{LAPS}}$ will shift the band edges of occupied (unoccupied) states on layer $i$ by $\Delta \varepsilon_{\mathrm{v},i}$ ($\Delta \varepsilon_{\mathrm{c},i}$).  

When the states of adjacent layers hybridize, it is not possible to find a basis of layer $i$ with occupation numbers exactly equal to 0 or 1. The best approximation to such a set of states is obtained by diagonalizing the density 
matrix of the entire heterostructure, 
\begin{equation}\label{eq:density}
\hat \rho = \sum_m^{\mathrm{occ}} |\psi_m\rangle \langle \psi_m|
\end{equation}
in a basis spanning the Hilbert space of layer $i$. Assuming that $\{\tilde \phi_{ni}\}$ represents such a basis, we define $|\rho_{ni}\rangle$ by diagonalizing the layer projected density matrix 
\begin{equation}
\rho_{\nu\mu}^i = \langle \tilde \phi_{\nu i} |\hat \rho |\tilde \phi_{\mu i}\rangle.
\end{equation}
The atom-centered LCAO basis functions implemented in GPAW could in principle qualify as the layer basis functions, except that they do fullfill the orthogonality criterion. We therefore use the orthonormal Löwdin transformed basis, 
\begin{equation}
| \tilde \phi_{\mu i} \rangle = \sum_{\nu,j}S^{-1/2}_{i\mu ,j\nu} | \phi_{\nu j} \rangle,  
\end{equation}
where $S$ is the LCAO overlap matrix. More details on the technical implementation of these equations are provided in Appendix~\ref{implementation}.

Although the LAPS operator should be included in the self-consistency loop, it can also be treated in a one-shot manner. In this case, the LAPS operator is evaluated from the density matrix of a self-consistent DFT calculation and the Kohn-Sham Hamiltonian is subsequently re-diagonalized including the LAPS operator. In the case of self-consistent LAPS, the self-consistency enters in two places. First, since the LAPS operator is added to the DFT Hamiltonian, the original wave functions and electron density are no longer self-consistent solutions of the Kohn-Sham equations and must be recomputed. Secondly, the LAPS operator itself depends on the density matrix, which changes during the self-consistency cycle.

\subsection{Determining the scissors shifts}
In the most general situation, the scissors shifts applied on layer $i$ contain three terms: 
\begin{eqnarray}\label{eq:E_val}
\Delta \varepsilon_{\mathrm{v},i} &=& \Delta \varepsilon^0_{\mathrm{vbm},i}+ \Delta \varepsilon_{\mathrm{vbm},i}^{\mathrm{strain}}+ \Delta \varepsilon_{\mathrm{vbm},i}^{\mathrm{ic}} \\\label{eq:E_con}
\Delta \varepsilon_{\mathrm{c},i} &=&\Delta \varepsilon^0_{\mathrm{cbm},i} + \Delta \varepsilon_{\mathrm{cbm},i}^{\mathrm{strain}}+ \Delta \varepsilon_{\mathrm{cbm},i}^{\mathrm{ic}}
\end{eqnarray}

The first terms in Eqs. \eqref{eq:E_val} and \eqref{eq:E_con} correct the Kohn-Sham band edges to the target QP energies of the \emph{freestanding} monolayer $i$. Thus
\begin{eqnarray}
\label{eq:E0_val}
    \Delta \varepsilon^0_{\mathrm{vbm},i} &=& \varepsilon_{\mathrm{vbm},i}^{\mathrm{QP}} -\varepsilon_{\mathrm{vbm},i}^{\mathrm{KS}} \\
      \Delta \varepsilon^0_{\mathrm{cbm},i} &=& \varepsilon_{\mathrm{cbm},i}^{\mathrm{QP}} -\varepsilon_{\mathrm{cbm},i}^{\mathrm{KS}}
\end{eqnarray}
The second terms correct for changes in the band edge energies due to undesired strain of layer $i$ stemming from the use of a size-restricted supercell.  The last terms account for the weakening of the screened Coulomb interaction in layer $i$ produced by the dielectric environment of layer $i$ (i.e. the other layer(s) of the heterostructure and/or a substrate). The weakening of the screened interaction leads to a renormalization of the QP energies, which can be understood as an image charge effect\cite{winther2017band,neaton2006renormalization,garcia2009polarization}.

In this work, the target QP band edge energies of the freestanding monolayers are taken from G$_0$W$_0$ performed on top of a PBE calculation.   
The strain correction is obtained as the difference between the PBE band edge energies of the fully relaxed monolayer and the monolayer in the heterostructure supercell. This correction should be included depending on whether the purpose is to simulate the actual heterostructure or a (potentially) more realistic configuration, where the strain in the individual monolayers has been relaxed.  
Finally, the image charge correction is calculated using the G$\Delta$W-QEH method\cite{winther2017band,andersen2015dielectric,gjerding2020efficient}. 

Because the image charge interaction is attractive, it always holds that $\Delta \varepsilon_{\mathrm{vbm},i}^{\mathrm{ic}}>0$ and $\Delta \varepsilon_{\mathrm{cbm},i}^{\mathrm{ic}}<0$, and hence, the band gap is always reduced. While, the magnitude of the two terms tends to be similar, their specific values depend on the particular heterostructure. 
Naturally, larger corrections occur for thicker heterostructures and/or structures on a substrate, because such configurations provide more screening. 
However, the image charge corrections also depend on the properties of the layer itself, see Ref.
\cite{winther2017band}.

\subsection{Computational details} 
All calculations were performed with the GPAW electronic structure code~\cite{enkovaara2010electronic,mortensen2024gpaw}. 
All the unit cells and atomic structures were relaxed 
using the PBE-D3 exchange-correlation (xc-) functional and a plane wave basis set with uniform 2D k-point grids of density 8/\AA$^{-1}$. 
The DFT and LAPS band structures were calculated using the PBE xc-functional and a basis set consisting of linear combinations of atomic orbitals (LCAO) of the double zeta polarized (DZP) type\cite{larsen2009localized}.
The G$_0$W$_0$ calculations were performed using 
PBE and norm-conserving PAW potentials as starting point, with plane wave basis with an 800 eV cutoff for the ground state. For 
the dielectric matrix and self-energy a cutoff extrapolation was used with a maximum value of 200 eV, on a grid of $12\times 12 \times 12$ k-points for bulk MoS$_2$ and $18\times 18$ for monolayers and slabs. For the 2D systems, a truncated Coulomb interaction was used to avoid interactions between periodically repeated layers. The $q=0$ divergence was treated using a semi-analytical model for the screened interaction\cite{rasmussen2016efficient}.  Spin-orbit coupling was included non-perturbatively unless stated otherwise. 

\section{Results and discussion}
 We begin by a study of multilayer MoS$_2$ as an example of a system without interfacial dipoles or in-plane strain effects. The full band structures obtained with LAPS and G$_0$W$_0$ are compared for the monolayer, bilayer and, bulk systems, while we further use LAPS to compute the band structure of stacks with up to 20 layers. After this example, we briefly review the problem of band alignment at semiconductor heterointerfaces from a general point of view. Next, we explore how the band alignment in bilayer MoS$_2$ evolves when the LAPS operator is used to shift all the bands in one layer relative to those in the other layer, thus mimicking the effect of a perpendicular electric field. This example is useful for illustrating some general concepts, including the role of self-consistency. Next, we consider the lattice matched heterobilayers MoS$_2$/WS$_2$ and MoSe$_2$/WSe$_2$. For these heterobilayers we cannot rely on G$_0$W$_0$ to describe charge transfer, and thus self-consistency becomes important. Therefore, we focus on illustrating the quantitative differences between PBE, HSE06, and LAPS. Finally, we use the LAPS operator with monolayer energies from G$_0$W$_0$ to calculate the band structure of the lattice mismatched MoSe$_2$/WS$_2$ heterostructure. For ten different low-strain supercells containing between 75 and 456 atoms, we obtain a direct interlayer band gap of 1.71-1.73 eV, in excellent agreement with experiments.


\subsection{MoS$_2$ multilayers: LAPS vs. G$_0$W$_0$}\label{sec:mos2}
%
As previously mentioned, the accuracy of the widely used one-shot G$_0$W$_0$@DFT method is questionable for general vdW heterostructures, because the calculation inherits the interfacial dipole built into the DFT starting point, which could be wrong. Obviously, this problem is not present for homo-multilayer structures, because they contain no interfacial dipoles. Moreover, such structures are perfectly lattice matched and can be modeled using minimal unit cells. G$_0$W$_0$ is therefore expected to constitute a reliable method for such systems (subject, of course, to its known limitations regarding lack of self-consistency and vertex corrections). 

\begin{figure*}
   \centering
    \includegraphics[width=0.99\textwidth]{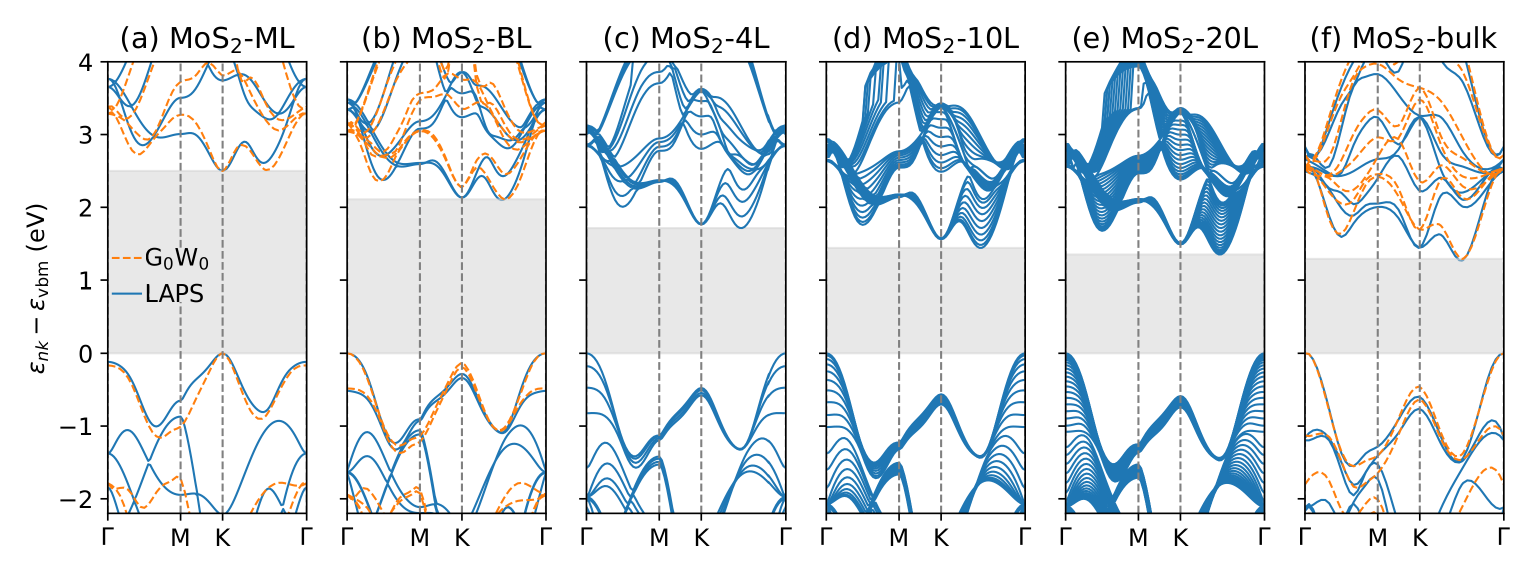}
    \caption{Band structure of MoS$_2$ monolayer, bilayer, 4-layer, 10-layer, 20-layer, and bulk. The band structures are calculated using G$_0$W$_0$ (dashed orange) and the LAPS operator (solid blue) with scissors shifts from a G$_0$W$_0$ monolayer calculation plus an image charge correction from the dielectric environment in the cases of the multilayer systems (see Table~\ref{tab:corrections}). Spin-orbit coupling is not included.}
    \label{fig:GWvsLAPS}
\end{figure*}

Fig. \ref{fig:GWvsLAPS} shows the band structure of a series of MoS$_2$ slabs, from monolayer to bulk. For all multilayer structures, the most stable AB-stacking order is used. The band structures calculated with LAPS are shown in blue. For monolayer, bilayer, and bulk we also show the G$_0$W$_0$ bands (orange color). The scissors shifts of the LAPS operator are provided in Table \ref{tab:corrections}. The correction $\Delta \varepsilon_n^0$ is obtained as the difference between PBE and G$_0$W$_0$ band edges of the isolated monolayer. This correction is the same for all the slabs.
The image-charge corrections, $\Delta \varepsilon_n^{\mathrm{ic}}$, for a given layer depends on the layer position in the slab (for slabs with more than two layers).
 In general, the correction is larger for the layers closer to the center of the slab, because the dielectric screening from the surrounding layers will be stronger. 

 By construction, the G$_0$W$_0$ band edge energies and band gap of the monolayer are exactly reproduced by LAPS. In the bilayer, the effects of interlayer hybridization and dielectric screening (image charge effect) reduce the band gap by 0.39 eV, from 2.50 eV to 2.11 eV (G$_0$W$_0$). In comparison, LAPS yields a band gap of 2.10 eV in excellent agreement with the target G$_0$W$_0$ value. In the bulk, the gap is reduced further to 1.29 eV (G$_0$W$_0$), which is also well reproduced by LAPS (1.27 eV). Both LAPS and G$_0$W$_0$ predict a direct band gap in the monolayer and indirect band gaps in the bilayer and bulk. We emphasize that all the LAPS calculations only require full G$_0$W$_0$ data for the freestanding monolayer.

\begin{table}
\center
\begin{ruledtabular}
\begin{tabular}{cccc}
  \\
   & $\Delta \varepsilon_i^0$ & $\Delta \varepsilon_i^{\mathrm{strain}}$ & $\Delta \varepsilon_i^{\mathrm{ic}}$ \\
  MoS$_2$ & vbm{ } cbm & vbm{ } cbm &  vbm{ } cbm \\ \hline \\[3pt]
  ML  & -0.20{ } 0.62 & 0.0{ } 0.0 & 0.0{ } 0.0 \\[2pt]
  BL  & -0.20{ } 0.62 & 0.0{ } 0.0 & 0.07{ } -0.09 \\[2pt]
  4L  & -0.20{ } 0.62 & 0.0{ } 0.0 & (0.12, 0.14){ } (-0.13, -0.16) \\[2pt]
  10L  & -0.20{ } 0.62 & 0.0{ } 0.0 & (0.15, 0.19){ } (-0.17, -0.21) \\[2pt]
  20L  & -0.20{ } 0.62 & 0.0{ } 0.0 & (0.17, 0.21){ } (-0.18, -0.23) \\[2pt]
  bulk  & -0.20{ } 0.62 & 0.0{ } 0.0 & 0.22{ } -0.23
     \\       
 \end{tabular}
 \end{ruledtabular}
 \caption{The scissors shifts used for the LAPS calculations of MoS$_2$ monolayer, bilayer, 4-layer, 10-layer, 20-layer, and bulk. For the 4, 10, and 20 layer systems the image charge corrections, $\Delta \varepsilon_i^{\mathrm{ic}}$, are layer dependent. For these systems the smallest and largest corrections are shown. The numerically larger (smaller) corrections are found for layers closest to the center (surface) of the slabs. All quantities are in eV.} 
  \label{tab:corrections}
 \end{table}

While the LAPS operator reproduces the G$_0$W$_0$ band edge energies accurately, it is clear that there are qualitative differences between the LAPS and G$_0$W$_0$ band structures, in particular for the second (third) valence band of the monolayer (bilayer). We stress that our G$_0$W$_0$ results are in good agreement with previous results in the literature\cite{molina2015vibrational}. We also note that concerning the qualitative shape of the band structure (i.e. ignoring the band gap size) the self-consistent GW$_0$ method yields results for both the monolayer and bilayer\cite{shi2013quasiparticle,cheiwchanchamnangij2012quasiparticle} that are somewhat intermediate between the LAPS and G$_0$W$_0$ results.


The high efficiency of LAPS in terms of computational cost makes it possible to obtain self-consistently corrected QP band structures of slabs with thickness much beyond what is feasible with GW.
The middle panels of Figure \ref{fig:GWvsLAPS} show the LAPS band structure of MoS$_2$ stacks with 4, 10, and 20 layers.  The excellent agreement between LAPS and G$_0$W$_0$ found for the band gap of monolayer, bilayer, and bulk, is expected to carry over to these multilayer structures. We note in passing that LAPS calculations could in principle be performed for MoS$_2$ slabs with hundreds of layers with modest efforts.     

\begin{figure*}
\centering
\includegraphics[width=1.0\textwidth]{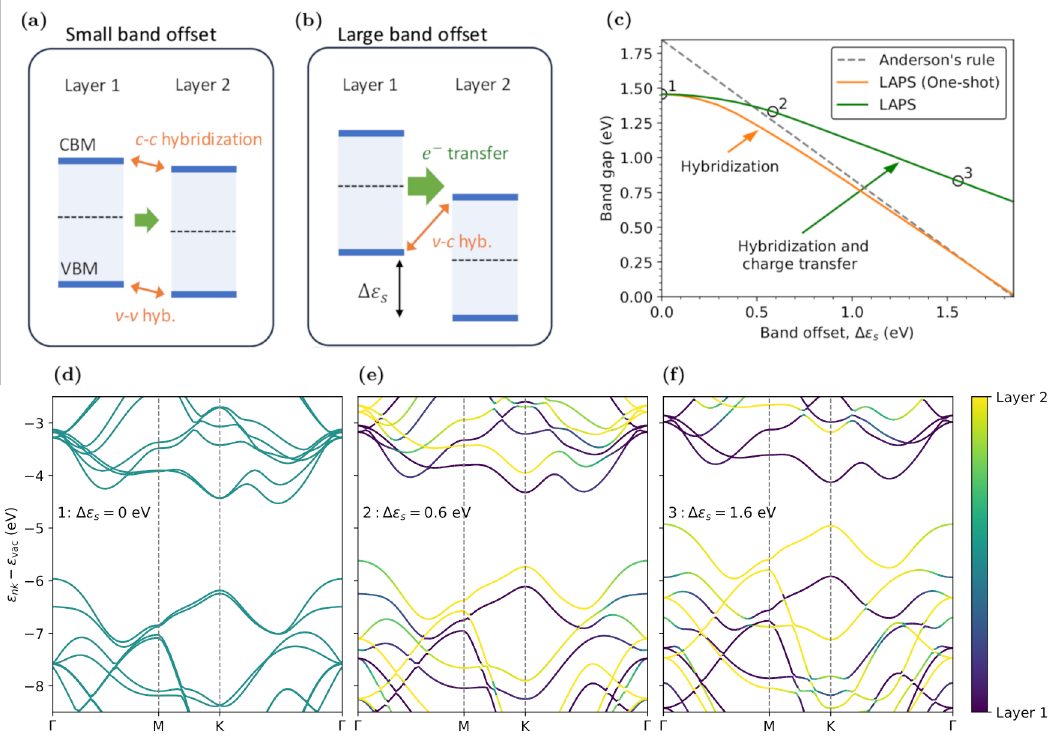}
\caption{(a, b) Schematic illustration of the factors governing band alignment at a semiconductor heterointerface. In case of a small band offset (a), the conduction and valence states will hybridize relatively strongly, while charge transfer effects will be relatively weak. The situation is opposite for the case of large band offsets (b).  
(c) Band gap of a MoS$_2$ homobilayer (AB stacked) as a function of the band offset, $\Delta \varepsilon_s$. The band offset is simulated by a LAPS operator shifting the bands (both occupied and unoccupied) of one of the layers by $\Delta \varepsilon_s$. The dashed line indicates the Anderson's rule\cite{anderson1960germanium}, which aligns the vacuum levels of the two layers and corresponds to a complete neglect of hybridization and charge transfer. The orange and green curves represent the result of the LAPS operator applied self-consistently and non-selfconsistently, respectively. (d-f) Full band structure of the MoS$_2$ bilayer calculated using the self-consistent LAPS method for the three values of $\Delta \varepsilon_s$ indicated in panel (c). Spin-orbit interactions are not included. The color code indicates the localization of the wave function on the two layers. 
}
\label{fig:sketch}
\end{figure*}

\subsection{Basics of band alignment}\label{sec:basics}
We consider two distinct monolayers, both of which are assumed to be non-metallic in their freestanding form. When the layers are far apart (dissociative limit), they will share a common vacuum level defined by the asymptotic value of the electrostatic potential in the region between the layers. This condition of a common vacuum level determines the relative alignment of the VBM and CBM of the two layers in the dissociative limit. This is referred to as Anderson's rule of band alignment\cite{anderson1960germanium}. 

Based on Anderson's rule, one can categorize the band alignment in three types: straddling gap (referred to as type I), staggered gap (type II), and broken gap (type III). Although the Anderson rule neglects interlayer interactions, it is expected to predict the type of band alignment well in vdW heterostructures, due to the relatively weak coupling between the layers. A previous work used the Anderson model to predict the band alignment type in thousand of vdW heterobilayers based on the band structure of 250 isolated monolayers calculated with PBE, HSE06, and G$_0$W$_0$, respectively. It was found that the band alignment type predicted by PBE and HSE06 was in disagreement with G$_0$W$_0$ for 44\% and 21\% of the bilayers, respectively\cite{haastrup2018computational}. 
This result clearly shows that the shortcomings of DFT-based band structures for vdW heterostructures are not only of quantitative nature (as known for homogeneous semiconductors), but can also be of qualitative nature.

We now consider effects beyond Anderson's rule. When the two layers are brought in proximity to each other, their wave functions will begin to hybridize. We distinguish between hybridization of bands with the same occupation (c-c and v-v) and opposite occupation (v-c and c-v, from hereon collectively referred to as v-c hybridization). From basic perturbation theory, it follows that hybridization is stronger between states closer in energy. The c-c and v-v hybridization tends to reduce the gap being more significant when the corresponding bands are more aligned (prior to hybridization), see the sketch in Fig.~\ref{fig:sketch}(a,b). Because the c-c and v-v hybridization mixes states of the same occupation, it will not result in a net charge transfer between the layers. 

The v-c hybridization also affects the band energies directly, but in contrast to the c-c/v-v hybridization,  it will \emph{increase} the gap. As a secondary effect, the v-c hybridization gives rise to charge transfer between the layers (or at least a charge redistribution). This happens because the hybridization mixes states of opposite occupation across the interface. In a self-consistent calculation, the charge redistribution and associated interface dipole will in turn influence the band alignment. Because the v-c hybridization produces dipoles in opposite directions, the net charge transfer will be larger for more asymmetric band alignments, i.e. larger band offset, see
Fig.~\ref{fig:sketch}(b). Just like the direct effect of the c-v and v-c hybridization, the interface dipole will also tend to increase the band gap. 

The concepts of hybridization and charge transfer are illustrated and discussed further in the following section. 

\subsection{MoS$_2$ bilayer in electric field: Importance of self-consistency}
In section \ref{sec:mos2}, we studied multilayer MoS$_2$ structures. For such homogeneous structures without any intrinsic band offset between the individual layers, the effect of charge transfer and interfacial dipoles, which are driven by asymmetric band alignments, will be very small (they may not vanish completely due to weak asymmetries in the band alignment caused by surface effects).  

To illustrate the effect of interfacial dipoles, as well as the role of self-consistency, we now consider homobilayer MoS$_2$ in the presence of a perpendicular electric field. The electric field will shift the average potential in the two layers leading to a type II band alignment. The effect of the electric field is simulated by a LAPS operator shifting all the bands in one of the two layers by a constant, $\Delta \varepsilon_s$.

 Fig.~\ref{fig:sketch}(c) shows the band gap of the MoS$_2$ homobilayer as a function of the band offset, $\Delta \varepsilon_s$. As these calculations are only for illustration purposes, spin-orbit interactions are not included. The dashed grey line shows the Anderson's rule, where the band edges of the monolayers are aligned with a common vacuum level. This result thus corresponds to neglecting both hybridization and charge transfer. Including the LAPS operator in a one-shot manner (orange line) accounts for hybridization but not charge transfer. As can be seen, the hybridization tends to reduce the band gap relative to the Anderson limit. This effect is due to v-v and c-c hybridization (see illustration in Fig.~\ref{fig:sketch}(a,b)). As the scissors shift is increased, the bands on the two layers are detuned in energy and the effect of v-v and c-c hybridization on the band energies diminishes. By including the LAPS operator self-consistently (green line) both hybridization and charge transfer effects are accounted for. The effect of charge transfer is to create an interface dipole that counteracts the scissors shift and increases the gap relative to the one-shot result. This effect becomes larger as the band offset is increased as the v-c hybridization is enhanced (see illustration in Fig.~\ref{fig:sketch}(a+b)). 

Fig.~\ref{fig:sketch}(d-f) shows the full band structure calculated with the LAPS operator for three different values of $\Delta \varepsilon_s$ also indicated by three circles in Fig.~\ref{fig:sketch}(c). The color of the bands reflects the localization of the corresponding wave function on the two layers. For $\Delta \varepsilon_s=0$ all bands are equally distributed on the two layers. As the values of $\Delta \varepsilon_s$ become larger, compared to the interlayer hybridization energy, the states become more and more localized on one of the layers.

\begin{figure}
    \centering
    \includegraphics[width=0.5\textwidth]{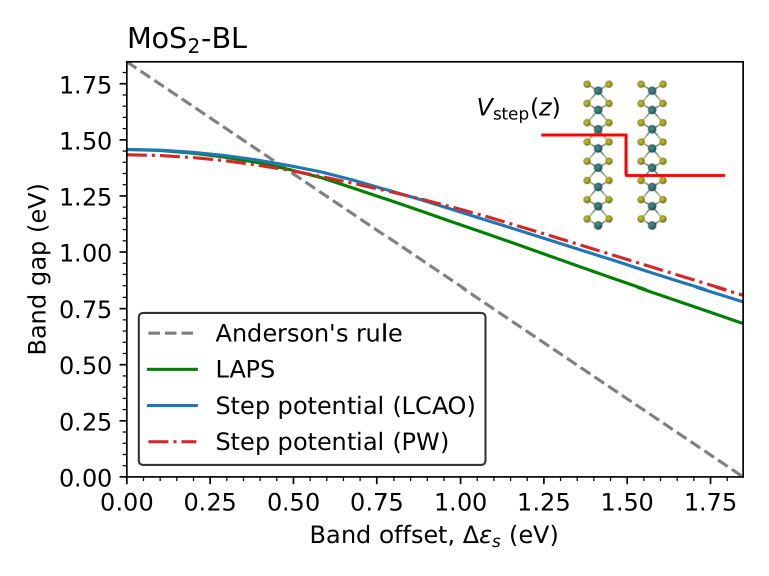} 
    \caption{Band gap of homobilayer MoS$_2$ as a function of the band offset. The latter is simulated in two different ways: By means of a LAPS operator as in Fig. \ref{fig:sketch}(c) (green) and by a step potential. The latter is calculated using both an LCAO basis set (blue) and plane wave basis set (red).}
    \label{fig:step}
\end{figure}

Above we used the LAPS operator to rigidly shift all the bands in one of the MoS$_2$ layers. Such an effect could also be simulated using a simple step function potential. In Fig.~\ref{fig:step} we compare the band gap of the MoS$_2$ bilayer as a function of the band offset included either by a LAPS operator (green - as in Fig. 1) or a step potential (blue). It is reassuring that the two approaches yield similar band gaps despite their very different technical implementation. Importantly, however, the LAPS operator is more general than the step function potential, as the latter can only shift all bands by the same amount while the former allows for occupation number-dependent shifts. We also note that the band gap result for the step potential is calculated with the LCAO-DZP basis (blue) and a plane wave basis (red) are in very close agreement.

\subsection{MoS$_2$/WS$_2$: LAPS vs. DFT}

\begin{figure}
    \centering
    \includegraphics[width=0.5\textwidth]{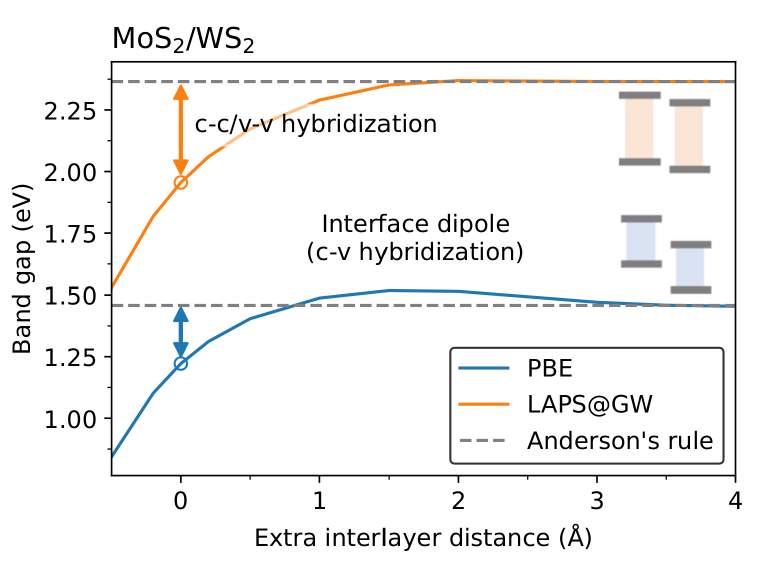}
    \caption{The band gap of the lattice matched vdW heterobilayer MoS$_2$/WS$_2$ as a function of the interlayer distance. The band gaps are calculated using DFT-PBE (blue) and a LAPS operator with target band edge energies from G$_0$W$_0$ (orange). In both cases the band gaps converge to the result predicted by the Anderson's rule. The band gap reduction at the equilibrium distance, due to v-v and c-c band hybridization, is indicated by vertical arrows. Band energies include spin-orbit interactions.}
    \label{fig:LAPSvsDFT}
\end{figure}

In the previous two sections, we explored some qualitative properties of the LAPS operator. We now consider its quantitative performance relative to a pure DFT description. Fig.~\ref{fig:LAPSvsDFT} shows the band gap of the lattice matched heterobilayer MoS$_2$/WS$_2$ (AB stacking) calculated with DFT-PBE and the LAPS operator, respectively. The target band edges used for the latter were obtained from G$_0$W$_0$ calculations.  All values include spin-orbit coupling. 

The first thing to notice is that PBE and LAPS yield rather different band gaps at the equilibrium distance. This is not surprising given that PBE underestimates the band gap of the monolayers by almost 1~eV.\cite{haastrup2018computational} It can be further seen that the Anderson rule overestimates the band gap at the equilibrium distance by 0.24 eV when PBE is used both for the bilayer and for the vacuum energies and 0.41 eV using LAPS. This significant difference is due to c-c and v-v hybridization. The fact that the effect of c-c/v-v hybridization is larger with LAPS than with PBE indicates that the (uncoupled) conduction and valence bands of the two monolayers are more aligned in G$_0$W$_0$ than in PBE (as illustrated by the band alignment sketches next to the two curves in Fig.~\ref{fig:LAPSvsDFT}). By comparing the band edges of the isolated monolayers we have verified that this is indeed the case (the relevant energies can be found, e.g. on the C2DB website\footnote{https://c2db.fysik.dtu.dk}). 

It can be further seen that the PBE gap overshoots the Anderson result for interlayer separations in the range of 1-3 \AA beyond the equilibrium value. This is due to v-c hybridization and the associated interface dipole. The fact that this effect does not occur in the LAPS calculation, is consistent with the PBE (LAPS) band alignment being more asymmetric (symmetric).

\subsection{Emulating hybrid DFT with LAPS}
The LAPS operator is intended to simulate many-body QP band structure of vdW heterostructures. However, it can of course be used to simulate band structures at any level of theory. For example, we can use the LAPS operator to simulate hybrid-functional DFT band structures at the cost of conventional DFT. Since hybrid DFT is much more affordable than self-consistent many-body approaches like GW, we can use self-consistent hybrid calculations to benchmark the accuracy of LAPS. 
For this purpose, we define the scissors shifts from the difference between the PBE and HSE06\cite{heyd2003hybrid} band structures of the isolated monolayers. We refer to this approach as LAPS@HSE.

\begin{figure}
    \centering
    \includegraphics[width=0.5\textwidth]{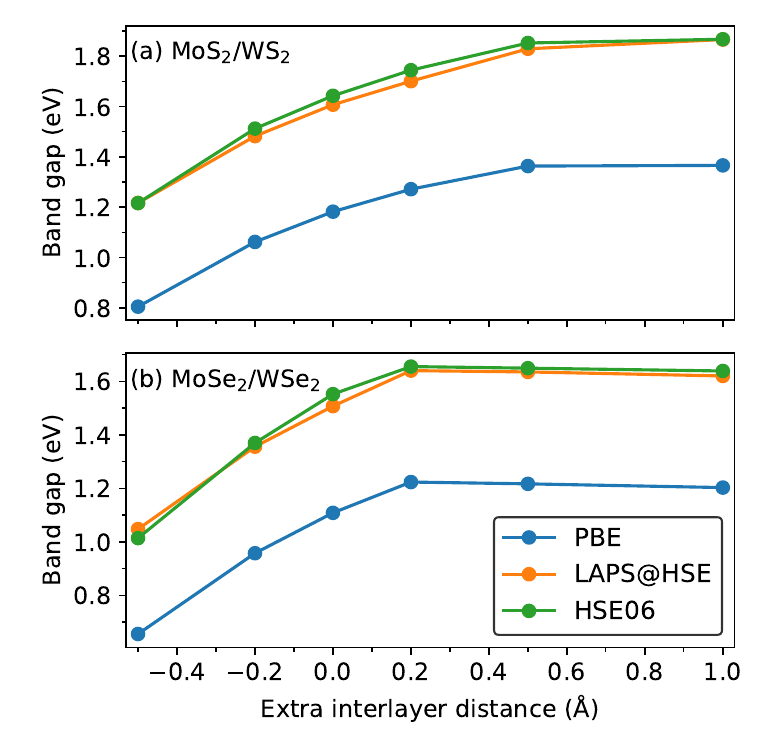}
    \caption{The band gap of the lattice matched vdW heterobilayer MoS$_2$/WS$_2$ (a) and MoSe$_2$/WSe$_2$ (b) as a function of the interlayer distance. The band gaps are calculated with DFT-PBE (blue), HSE06 (green), and a LAPS operator with target band edge energies from HSE06 (orange).}
    \label{fig:LAPSforHSE}
\end{figure}

Figure \ref{fig:LAPSforHSE}
shows the band gap of the lattice matched MoS$_2$/WS$_2$ and MoSe$_2$/WSe$_2$ heterobilayers calculated using pure PBE (blue), HSE06 (green), and LAPS@HSE (orange). 
All calculations are performed self-consistently without the inclusion of spin-orbit interactions. Across all the interlayer distances, LAPS@HSE yields band gaps in excellent agreement with the full HSE06 results for both bilayers. This shows that the interlayer interactions can be described at the PBE level (as done by LAPS), as long as the HSE06 scissors corrections are applied to the individual layers. In contrast, PBE without scissors corrections underestimates the gap by about 0.4 eV for both heterobilayers. 

\begin{figure*}
\centering
\includegraphics[width=0.99\textwidth]{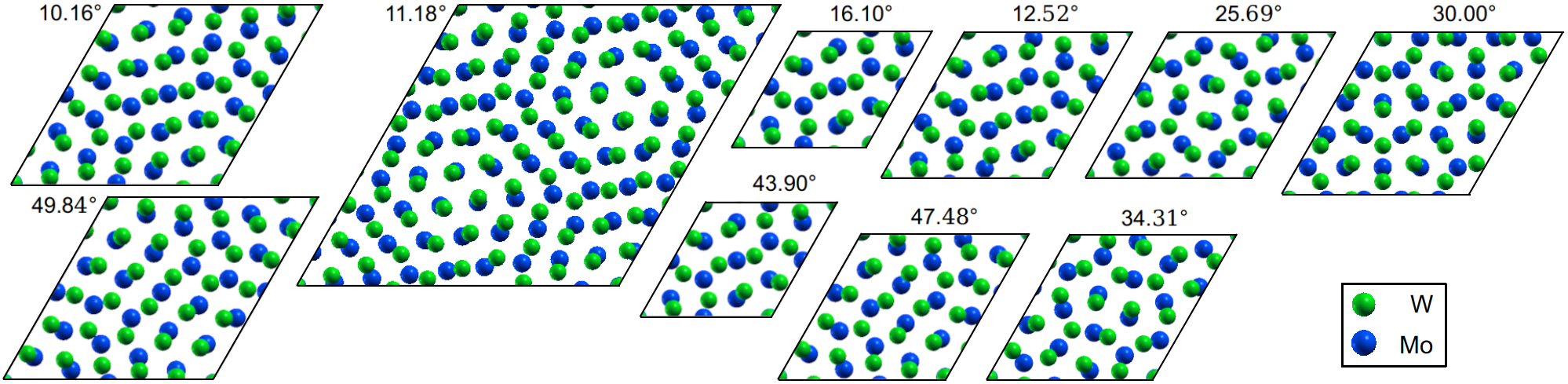}        
\caption{The moir\'e cells of the MoSe$_2$/WS$_2$ heterostructures considered in this work (only W and Mo atoms are shown). The relative twist angle of the two monolayers is indicated next to each cell.}
\label{fig:atoms456}
\end{figure*}

\subsection{MoSe$_2$/WS$_2$ moir\'e structures}
In this section we use the LAPS operator to calculate the band structure of the MoSe$_2$/WS$_2$ heterostructure. Because the lattice constants of the two monolayers differ significantly (by $\sim 4$\%), any stacking of the layers will result in a moir\'e structure with periodicity exceeding the primitive lattice of both monolayers. Model heterostructures are created by combining the unit cell basis vectors of the MoSe$_2$ and WS$_2$ monolayers to identify supercells (from hereon referred to as moir\'e cells) in which the two monolayers are strained by less than 1\%, see Ref. \onlinecite{sauer2025dispersion} for a detailed description of the method used to create the moir\'e cells. Because the heterobilayers are relaxed using PBE-D3 to account for the van der Waals interactions between the layers, we also use monolayer lattice vectors from PBE-D3 calculations when constructing the moir\'e cells. Table \ref{tab:angles} provides an overview of the identified moir\'e structures. The first three columns show the twist angle between the monolayers, the number of atoms in the moir\'e cell, and the maximum strain on the monolayers. As can be seen, the structures contain between 75 and 456 atoms in the moir\'e cell and the strain on each monolayer is below 0.55\% in all structures. The moir\'e cells of all the structures are depicted in Fig. \ref{fig:atoms456} (only W and Mo atoms are shown).  

\begin{table}
\center
\begin{ruledtabular}
\begin{tabular}{ccccc}
  \\[-5pt]
   $\alpha$ (deg) & $N_{\text{atoms}}$ & Strain (\%)  & $E_{\mathrm{gap}}^{\mathrm{LAPS}}$ (eV) &  $E_{\mathrm{gap}}^{\mathrm{LAPS}}$ (eV)  \\
   &&&& w. strain corr.\\
[5pt]
   \hline \\[3pt]         

  10.16 & 177 & 0.55 & 1.79& 1.71\\
  11.18 & 456 & 0.09 & 1.79& 1.72\\
  12.52 & 120 & 0.50 & 1.81& 1.71\\
  16.10 &  75 & 0.06 & 1.79& 1.73\\
  25.69 & 120 & 0.50 & 1.82& 1.71\\
  30.0 & 156 & 0.15 & 1.78& 1.73\\
  34.31 & 120 & 0.50 & 1.83& 1.72\\
  43.90 &  75 & 0.06 & 1.79& 1.72\\
  47.48 & 120 & 0.50 & 1.80& 1.71\\
  49.84 & 177 & 0.55 & 1.79& 1.71
 
   \\[5pt]
 \end{tabular}
 \end{ruledtabular}
 \caption{Overview of the MoSe$_2$/WS$_2$ moir\'e structures considered in this work. The first three columns show the twist angle ($\alpha$), number of atoms in the moir\'e cell ($N_{\mathrm{atoms}}$), and maximum strain component on the two monolayers. The last two columns show the band gap calculated with LAPS using the G$_0$W$_0$ band energies of the monolayers with image charge corrections. Results are shown both without (second-last) and with (last column) strain corrections included in the scissors shifts. All calculations include spin-orbit interactions.}  
  \label{tab:angles}
 \end{table}

\begin{figure}
\centering
\includegraphics[width=0.5\textwidth]{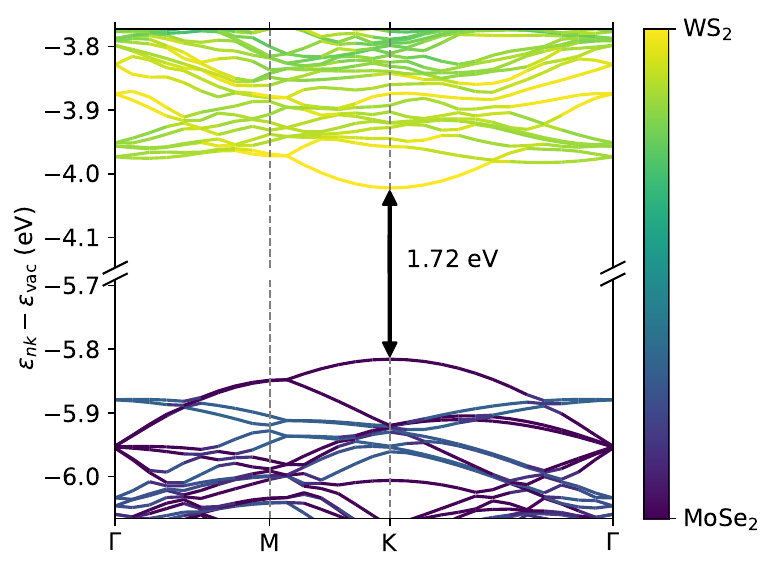}        
\caption{The LAPS@GW band structure of the MoSe$_2$/WS$_2$ bilayer with a twist angle of $11.18^{\circ}$ and 456 atoms in the moir\'e cell. Spin-orbit interactions are included and all energies are referenced to the vacuum level.}
\label{fig:bs456}
\end{figure}

\begin{figure}
\centering
\includegraphics[width=0.5\textwidth]{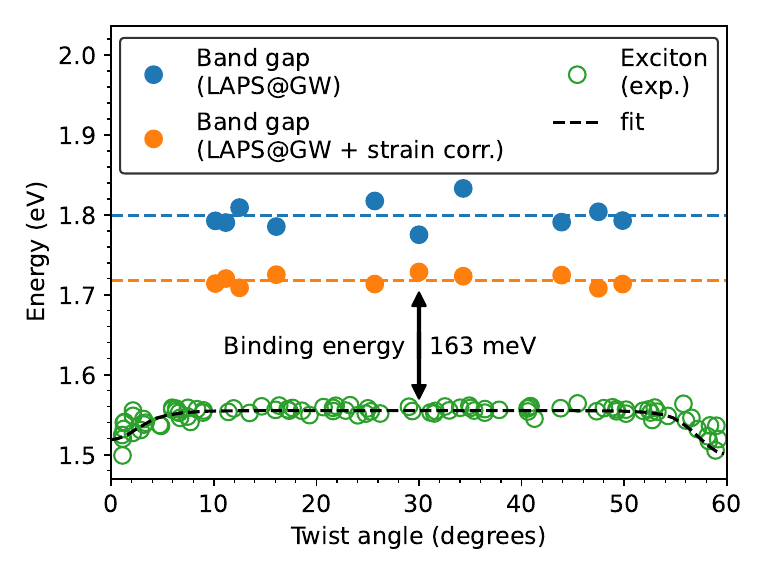}        
\caption{Experimental interlayer exciton energies (green circles)\cite{alexeev2019resonantly} and calculated QP band gap (blue and orange circles) of the MoSe$_2$/WS$_2$ heterobilayer as a function of relative twist angle. The band gaps are calculated using LAPS, with monolayer energies from G$_0$W$_0$, and image charge corrections from G$\Delta$W. Results are shown both with and without inclusion of strain corrections in the scissors shifts. The band gaps include spin-orbit interactions.}
\label{fig:exp-vs-LAPS}
\end{figure}

All the considered MoSe$_2$/WS$_2$ heterostructures exhibit a type II band alignment, see Fig. \ref{fig:bs456} for an example of a band structure. The (interlayer) band gaps calculated with LAPS as function of the twist angle are shown in Fig. \ref{fig:exp-vs-LAPS}. The scissors shifts are based on G$_0$W$_0$ results for the isolated monolayers and image charge corrections from G$\Delta$W, both of which are independent of the twist angle. The image charge corrections are given by  $(\varepsilon_v^{\mathrm{ic}},\varepsilon_c^{\mathrm{ic}})$=$(59,-72)$ meV for MoSe$_2$ and $(91,-100)$ meV for WS$_2$. The LAPS results are shown both with and without corrections for in-plane strain in the scissors shifts. The inclusion of the strain correction reduces the variation in the band gap with respect to twist angle, from 0.06 eV to 0.02 eV. For comparison, the experimentally determined position of the lowest (interlayer) exciton is also shown (green symbols)\cite{alexeev2019resonantly}. In general, the difference between the QP band gap and exciton energy amounts to the exciton binding energy. The strain-corrected LAPS band gap and the measured exciton energy differ by 163 meV. This is in fair agreement with experimental estimates and first-principles BSE calculations (for lattice matched TMD heterobilayers), which lie on the range 200-300 meV.
\cite{rivera2016valley,gillen2018interlayer,torun2018interlayer}.   
On this basis, we conclude that the band gaps predicted by G$_0$W$_0$-based LAPS
for the MoSe$_2$/WS$_2$ moir\'e structures, are very reasonable. In particular, when considering that the difference of $<150$ meV between the LAPS-derived values for the exciton binding energy and the literature references, is very comparable to the accuracy of the G$_0$W$_0$ for homogeneous semiconductors. Thus, it is likely that the deviation should be ascribed to the G$_0$W$_0$ monolayer target values method rather than the LAPS method. We note that direct measurement of the QP gap of vdW heterostructures is a significant challenge, and we have not found such data for MoSe$_2$/WS$_2$. 

\section{Conclusion}
We have introduced a layer-projected scissors (LAPS) operator that enables efficient calculations of single-particle band structures of general vdW heterostructures with beyond-DFT accuracy at the cost of conventional DFT. The LAPS operator is included in the self-consistent DFT calculation to correct the states on each layer by a layer- and occupation number-dependent scissors shift that mimics the many-body self-energy. The scissors shifts entering the LAPS operator contain three terms (for the occupied and unoccupied states on each layer): (1) The \emph{monolayer shift}, which corrects the DFT band edges of the isolated monolayer to match the 'target' band edge energies. The target values for the band edge energies of the monolayer could come from experiments or higher level calculations such as GW. (2) The \emph{image charge shift}, which accounts for the renormalization of the band energies in a given layer due to environmental screening coming from other layers or substrates. This shift is calculated using the QEH-G$\Delta$W method. (3) The \emph{strain shift}, which accounts for changes in the band edge energies of a layer due to unphysical strain effects resulting from the use of a model supercell. The strain shifts are calculated at the DFT level.  

Via its dependence on the density matrix, the LAPS operator changes during the self-consistency cycle (the size of the scissors shifts remain constant, though). This in turn affects the DFT effective potential and wave functions. We have found that the self-consistent treatment of the LAPS operator can be essential. This is true in particular, when the LAPS-induced changes in the band alignment affects the hybridization between occupied and unoccupied states across the vdW gap (v-c hybridization) and thereby induces interfacial dipoles.

We showed that the LAPS operator yields excellent results for multilayer MoS$_2$ structures with the target G$_0$W$_0$ band edge energies of bilayer and bulk being reproduced to within 0.02 eV. At the same time, the computational cost of LAPS is comparable to standard DFT making it applicable to very thick structures (here demonstrated for structures with up to 20 layers).  

Finally, we used the LAPS operator to calculate the twist-angle dependent QP band structure of MoSe$_2$/WS$_2$ heterobilayers containing up to 456 atoms in a moir\'e cell. The interlayer band gaps calculated on basis on G$_0$W$_0$ results for the isolated monolayers were found to lie 0.16 eV above the experimentally measured lowest interlayer exciton. This value corresponds well to the expected binding energy of a TMD interlayer exciton.

By enabling quantitatively accurate QP band structure calculations for 2D vdW structures with DFT efficiency, the LAPS operator opens the door to first-principles determination of a range of physical properties including optical spectra, carrier transport and dynamics, electron-phonon coupling, and topological invariants, which has so far been elusive for many experimentally relevant vdW materials.

\appendix
\section{Implementation of the LAPS operator}{\label{implementation}}

In the following, we derive the LCAO matrix representation of the scissors operator $\hat \Sigma_{\mathrm{LAPS}}$ (in the following just $\hat \Sigma$) as it is implemented in GPAW. We thus set out to determine the matrix 
\begin{equation}
\Sigma_{\mu \nu}=\langle \phi_{\mu} |\hat \Sigma|\phi_{\nu}\rangle
\end{equation}
where $\phi_\mu$ and $\phi_{\nu}$ are the (non-orthogonal) LCAO basis functions. With this matrix at hand we can solve the generalized Kohn-Sham eigenvalue problem 
\begin{equation}
\sum_\nu (H + \Sigma)_{\mu\nu} C_{\nu }^n
= \sum_{\nu} S_{\mu\nu} C_{\nu}^n \epsilon_n,
\end{equation}
where $S$ is the overlap matrix and $C_{\nu}^n$ the expansion coefficients of eigenstate $n$ in the LCAO basis.

Evaluating the matrix element in Eq. (3), the layer-projected density matrix takes the form
\begin{equation}
\rho^i_{\nu\mu} = \sum_n \sum_{k,j}^{\mathrm{layers}}\sum_{\nu'\mu'} S^{1/2}_{i\nu,k\nu'}C^n_{k\nu'}f_n C_{j\mu'}^{n*}S^{1/2}_{j\mu',i\mu},
\end{equation}
where $f_n$ is the occupation of eigenstate $n$ and we have used the basis function subscript $(i\nu)$ to indicate that basis function $\nu$ is centered on an atom in layer $i$. This matrix is diagonalized (with eigenvectors $D^m_{i\nu}$ and eigenvalues $\rho_{mi}$) yielding
\begin{equation}
\hat \rho^i = \sum_m \rho_{mi}|\rho_{mi}\rangle \langle \rho_{mi}|
\end{equation}
with 
\begin{equation}
|\rho_{mi}\rangle = \sum_{k}^{\mathrm{layers}}\sum_{\nu\mu} D^{m}_{i \nu} S^{-1/2}_{i\nu,k\mu} |\phi_{k\mu} \rangle.
\end{equation}
Writing $\hat \Sigma$ in the form 
\begin{equation}
\hat \Sigma = \sum_i^{\mathrm{layers}} \sum_m \Sigma_{mi} |\rho_{mi}\rangle \langle \rho_{mi}|,
\end{equation}
where $\Sigma_{mi} = \Delta \varepsilon_{\mathrm{v},i} \rho_{mi} + \Delta \varepsilon_{\mathrm{c},i} (1-\rho_{mi})$,
we obtain 
\begin{equation}
\Sigma_{k\nu,l\mu} = \sum_i^{\mathrm{layers}} \sum_m \sum_{\nu'\mu'} S^{1/2}_{k\nu,i\nu'} D^{mi}_{\nu'}\Sigma_{mi} D^{mi*}_{\mu'}S^{1/2}_{i\mu',l\mu}.
\end{equation}

\begin{acknowledgments}
K. S. T. is a Villum Investigator supported by VILLUM FONDEN (grant no. 37789).
Work by D. A. L. and K. B. was funded by the Research Council of Norway through the MORTY project (315330). All authors acknowledge support from the Novo Nordisk Foundation Data Science Research Infrastructure 2022 Grant: A high-performance computing infrastructure for data-driven research on sustainable energy materials, Grant no. NNF22OC0078009. M.K.S acknowledges support from the Novo Nordisk Foundation, Grant number NNF22SA0081175, NNF Quantum Computing Programme

\end{acknowledgments}

\newpage
\bibliography{paper}
\bibliographystyle{apsrev4-1}
\end{document}